\documentclass[12pt]{article}
\usepackage{amsmath,amssymb}
\usepackage{graphicx}   % for including figures
\usepackage[caption=false]{caption}    % for captions
\usepackage[font=footnotesize]{subfig} % subfig.sty for a double column floating figure using two subfigures
\usepackage{graphicx}   % for including figures
\usepackage[caption=false]{caption}    % for captions
\usepackage[font=footnotesize]{subfig} % subfig.sty for a double column floating figure using two subfigures
\usepackage{fixltx2e}
\usepackage{url}
\newcommand{\shorttitle}[1]%
{\markboth{Proceedings of the 31\MakeLowercase{$^{st}$} ICRC, {\L}\'{o}d\'{z} 2009}{#1} }
\newcommand{\etal}{\MakeLowercase{\textit{et al. }}} % "et al."
\begin{document}
\title{On the detectability of primordial black holes in the Galaxy}
\author{Julia K.\ Becker\\
{\normalsize Insitutionen f\"or Fysik, G\"oteborgs Universitet, 41296 G\"oteborg, Sweden}\\[0.5cm]
 Marek A.\ Abramowicz\\
{\normalsize Insitutionen f\"or Fysik,
  G\"oteborgs Universitet, 41296 G\"oteborg, Sweden}\\
{\normalsize N. Copernicus Astronomical Centre, }\\
{\normalsize Polish Academy of
  Sciences, Bartycka 18, 00-716 Warszawa, Poland}\\[0.5cm]
                          Peter
                          L.\ Biermann\\{\normalsize MPI for Radioastronomy, Bonn,
  Germany}\\{\normalsize Bonn University, Germany}\\{\normalsize Dept. of Phys. \& Astr., Univ. of Alabama,
Tuscaloosa, AL, USA}\\{\normalsize Dept. of Phys., Univ. of Alabama at Huntsville,
AL, USA}\\{\normalsize FZ Karlsruhe, and Phys. Dept., Univ. Karlsruhe, Germany}}
\shorttitle{Becker \etal Primordial black holes in the Galaxy}
\maketitle

\begin{abstract}
In the mass range of $10^{15}$~g up to $10^{26}$~g, primordial black holes (PBHs)
as a possible contribution to the dark matter are still unexplored. In
this contribution, we investigate the possibility of an
electromagnetic signal from PBH interactions with astrophysical
objects in the Galaxy. We find that a signal from passages cannot be
observed, since, depending on the mass, either the interaction probability or the energy loss
is too small. Further, we discuss possible effects from
high-mass PBHs at masses $>10^{26}$~g, where PBHs can still contribute to
the dark matter at the order of $\sim 10\%$. Here, we find that a
significant fraction of PBHs can be captured in the Hubble time. These
captures could therefore lead to detectable effects.
\end{abstract}

%%%%%%%%%%%%%%%%%%%%%%%%%%%%%%%%%%%%%%%%
\section{Introduction}
\begin{figure*}[!t]
\centering
\includegraphics[width=12cm]{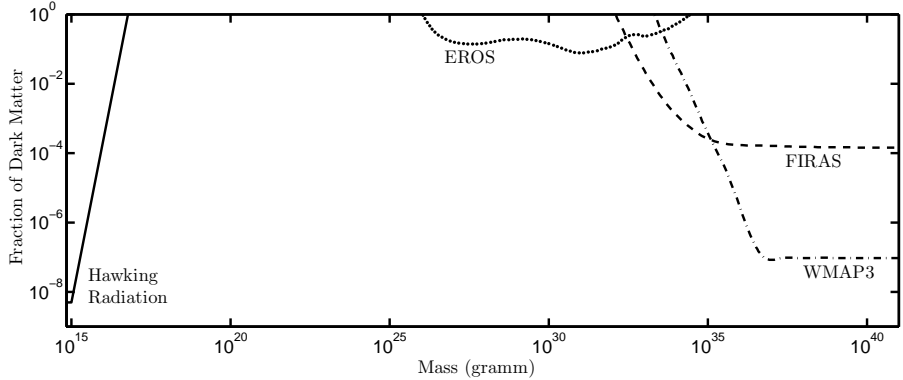}
\caption{Constraints on the contribution of PBHs to the dark matter}
\label{pbh_limits:fig}
\end{figure*}

%%%%%%%%%%%%%%%%%%%%%%%%%%%%%%%%%%%%%%%%
Although the existence of dark matter was already established by
Zwicky in 1933 \cite{zwicky1933}, its origin and nature is still unknown. Several
models were proposed to explain the missing matter, among others
weakly interacting massive particles (WIMPs),
sterile neutrinos, regular neutrinos, massive astrophysical compact
halo objects (MACHOS)
and also primordial black holes (PBHs). MACHOs and regular neutrinos have
been shown to have a negligible effect on dark matter. Primordial
black holes, on the other hand, can contribute up to 100\% in the
mass range $10^{15}$~g~$<m_{pbh}<10^{26}$~g. Here, $m_{pbh}$ is the
mass of the primordial black hole. Below and above this mass range, the contribution
of PBHs to the dark matter is constrained, with current limits shown
in Fig.\ \ref{pbh_limits:fig}: PBHs with masses at $5\cdot 10^{14}$~g
are expected to evaporate in the local universe, with a unique
signature of high-energy photon emission with energies above
$20$~MeV. Measurements by the EGRET experiment do not show any evidence of PBH evaporation
signatures and the contribution of PBHs to dark matter is strongly constrained
to a fraction of less than $3\cdot 10^{-9}$
\cite{egret,macgibbon_carr1991}. New data from the Fermi experiment
will provide further improvement to this limit. The region above
$10^{26}$~g is constrained to a level of less than $\sim 10\%$ of the
dark matter and for PBHs of the order of solar masses and above, constraints
are much more stringent, up to $10^{-8}$, see e.g.\ \cite{ricotti2008} and references therein.

In this paper, we investigate if it is possible to detect PBHs, or
alternatively to constrain their contribution to the dark matter, by
their interactions with stellar objects in the Galaxy. In order to do
this, we discuss the following questions: In Section
\ref{interaction:sec}, we discuss how often 
interactions between PBHs and stellar objects happen. We answer the
question of the energy loss in a single passage in Section
\ref{energy_loss:sec}. In Section \ref{captures:sec}, we investigate at what point is the PBH
captured by stellar objects and possible consequences. Section
\ref{conclusions:sec} discusses the results. All our results are based
on simple estimates, without considering exact distributions of the
PBH masses, velocities or the spatial distribution of dark
matter. Effects of including these effects are either negligible or
would decrease the sensitivity to the observation of PBH
interactions (see \cite{pbhs_apj2009} for a detailed discussion). The aim of this paper is to give an upper limit on
possible effects.
%%%%%%%%%%%%%%%%%%%%%%%%%%%%%%%%%%%%%%%%
\section{Interaction probability\label{interaction:sec}}
%%%%%%%%%%%%%%%%%%%%%%%%%%%%%%%%%%%%%%%%
In the following calculation, we consider the interaction of PBHs with
four classes of objects: main sequence stars (MS$\star$), red giant
cores (RGC), white dwarfs (WD) and neutron stars (N$\star$)
The rate of PBH interactions with a class of stellar objects is given by
\begin{equation}
\dot{n}\approx N_{\star}\cdot j_{pbh}\cdot \sigma_{pbh,\star}\,.
\end{equation}
Here, $N_{\star}$ is the number of stellar objects in the considered
class. The flux of primordial black holes, $j_{pbh}$, is given by the
product of
the PBHs' number density $n_{pbh}$ with their velocity $v_0$:
\begin{equation}
j_{pbh}=n_{pbh}\cdot v_0\,.
\end{equation}
We use the measured rotational velocity of matter in the Galaxy as the
relative velocity between PBHs and stellar objects, $v_0\approx 2.2\cdot 10^{7}$~cm\,s$^{-1}$. The number density of primordial black holes
can be determined by assuming that PBHs represent a fraction $\eta$ of the DM
in the Galactic halo, $M_{DM}\approx 9\cdot 10^{11}\,M_{\odot}$.
The spherical halo has an
approximate radius of $r_{halo}\sim 50$~kpc. We further assume that all PBHs are
produced at the same mass value $m_{pbh}$, with $m_{pbh}$ as a free
parameter. Thus, the number density of PBHs is
\begin{equation}
n_{pbh}=\frac{\eta\cdot M_{DM}}{4/3\,\pi\,r_{halo}^{3}}\cdot \frac{1}{m_{pbh}}
\end{equation}
and the event rate can be written as
\begin{equation}
\dot{n}\approx N_{\star}\cdot \frac{\eta\cdot M_{DM}}{4/3\,\pi\,r_{halo}^{3}}\cdot \frac{v_0}{m_{pbh}}\cdot \sigma_{pbh,\star}\,.
\label{ndot:equ1}
\end{equation}
The cross section of a PBH interacting with a
star is given as
\begin{equation}
\sigma_{pbh\,\star}=\pi\,R_{0}^{2}\,.
\end{equation}
Here, we use the
conservation of angular momentum in order to estimate the impact parameter
$R_0$, which guarantees the
collision with the star. Angular momentum conservation gives the relation
\begin{equation}
v_0\,R_0\approx v_{\star}\cdot  R_{\star}
\label{ang_mom}
\end{equation}
with the PBH's
velocity $v_{\star}$ when it reaches the surface of the star. The
final velocity is determined using energy conservation:
\begin{equation}
\frac{1}{2}v_{0}^{2}-\frac{G\,M_{\star}}{R_{dist}}=\frac{1}{2}v_{\star}^2-\frac{G\,M_{\star}}{R_{\star}}
\end{equation}
with $M_{\star}$ as the mass of the stellar object.
The characteristic distance of the PBH entering the gravitational
field of the star can be approximated as the typical distance
between objects in the galaxy, $R_{dist}$, assuming that each star-like objects
deflects the PBH at least marginally. We assume this typical distance
to be $R_{dist}\sim 1$~pc.
The gravitational energy is negligible in the
initial state\footnote{For $R_{dist}=1$~pc, we get
  $G\,M_{\star}/R_{dist}\approx 7000$~cm\,s$^{-1}$} and the final velocity can be expressed as
\begin{equation}
v_{\star}\approx
\sqrt{v_{0}^{2}+\frac{2\,G\,M_{\star}}{R_{\star}}}=\sqrt{v_{0}^{2}+v_{esc}^{2}}\,.
\label{vstar}
\end{equation}
Using Equ.~(\ref{vstar}) in Equ.~(\ref{ang_mom}) results in the correlation
\begin{equation}
R_{0}=R_{\star}\cdot \frac{v_{*}}{v_0}=R_{\star}\cdot
\frac{\sqrt{v_{0}^{2}+v_{esc}^{2}}}{v_0}\,.
\label{impact_parameter}
\end{equation}
Replacing the impact parameter $R_0$ in Equ.~(\ref{ndot:equ1}) by
using Equ.~(\ref{impact_parameter}) yields
\begin{eqnarray}
\dot{n}&\approx& N_{\star}\cdot \frac{\eta\cdot M_{DM}}{4/3\,r_{halo}^{3}}\cdot
\frac{v_0}{m_{pbh}}\cdot \left(\frac{v_{\star}}{v_0}\right)^{2}\cdot R_{\star}^{2}\\
&=&2\cdot 10^{-3}{\rm yr}^{-1}\cdot N_{\star,9}\cdot
\eta_{1}
\cdot\nonumber\\
&\quad&v_{\star,0}^2\cdot R_{\star,9}^{2}\cdot
m_{pbh,20}^{-1}\,.
\label{ndot:equ2}
\end{eqnarray}
We use fixed numbers for the dark matter mass in the Galaxy,
$M_{DM}=9\cdot 10^{11}\,M_{\odot}$ and for the halo radius,
$r_{halo}=50$~kpc. Other variables are
$N_{\star,9}:=N_{\star}/10^{9}$,
$\eta_{1}:=\eta/1$, $v_{\star,0}:=v_{\star}/v_0$,
$R_{\star,9}:=R_{\star}/(10^{9}{\rm  cm})$ and
$m_{pbh,20}/(10^{20}\,{\rm g})$.

Apart from the PBH mass, which we leave as a variable, the remaining parameters depend on the property of the
astrophysical objects considered. The basic properties of the source
classes are listed in table \ref{source_class_properties:tab}. The
resulting event rate scales as $m_{pbh,20}^{-1}$ and the values for
$\dot{n}(m_{pbh,20}=1)$ are given in table \ref{results:tab}.
  \begin{table}[!h]
  \caption{Properties of source classes in the Galaxy, considered as
    candidates for interactions with PBHs.}
  \label{source_class_properties:tab}
  \centering
  \begin{tabular}{|l|l|l|l|l|}
  \hline
   parameter &MS$\star$  & RGC & WD & N$\star$\\
   \hline 
   $N_{\star}$&$10^{11}$&$10^{9}$&$10^{9}$&$10^{9}$ \\
   $M_{\star}/M_{\odot}$&$1$&$1$ &$1$ &$1$  \\
   $R_{\star}$ [cm]&$10^{11}$&$10^{10}$ &$10^{9}$ &$10^{6}$ \\
   $\rho_{\star}$ [g cm$^{-3}$]&$10^{2}$&$10^{4}$&$10^{5}$&$10^{13}$ \\
  \hline
  \end{tabular}
  \end{table}
%%%%%%%%%%%%%%%%%%%%%%%%%%%%%%%%%%%%%%%%
\section{Energy loss in a single passage\label{energy_loss:sec}}
%%%%%%%%%%%%%%%%%%%%%%%%%%%%%%%%%%%%%%%%
In this estimate, we follow the calculations of Ruderman \& Spiegel
\cite{ruderman_spiegel1971}, originally done for a galaxy moving
through the interstellar medium. This logic can directly be
transferred to the problem of a PBH moving through a stellar object. Here, it is assumed that the kinetic energy which is transferred from the
black hole to the star is radiated. In this first estimate, we assume a uniform and
frictionless medium and we disregard the fact that the gas is
self-gravitating. For more details, see \cite{pbhs_apj2009}.

In first order approximation, we consider the motion of the black hole to be at constant
speed. 
In addition, we
 assume the velocity to be supersonic, so that the drag force, and hence for the energy loss, reduces to the one presented
in~\cite{ruderman_spiegel1971}. The bow shock is negligible
compared to the tail shock's energy loss.
Consequently, the energy loss per time is constrained to
\begin{equation}
\frac{dE_{\rm loss}}{dt}=\frac{4 \pi G^2 m_{pbh}^{2} \rho_{\star}}{v_{{\star}}} \ln{\frac{r_{\max}}{r_{\min}}}\,,
\label{eq:power_loss}
\end{equation}
where $r_{\max}$ is the maximal linear
dimension of the star (i.e.~its diameter),
$r_{\min}$ is the linear dimension of the accretor, lying between the black
hole horizon and the Bondi-Hoyle radius of the black hole.
Here, we use
$\ln(r_{\max}/r_{\min})=10$ as an estimate of the logarithmic ratio.
The total energy loss is then determined to
\begin{equation}
E_{\rm loss}=\frac{dE_{\rm loss}}{dt}\cdot dt=\frac{8 \pi G^2 m_{pbh}^{2} R_{\star} \rho_{\star}}{v_{{\star}}^{2}} \ln{\frac{r_{\max}}{r_{\min}}}\,,
\label{eq:en_loss}
\end{equation}
at a passage time of $dt\approx 2\cdot
R_{{\star}}/v_{{\star}}$. Then, Equ.~(\ref{eq:en_loss}) can be expressed
as
\begin{equation}
E_{\rm loss}=2\cdot 10^{27}\,{\rm erg}\,\cdot m_{pbh,20}^{2}\cdot R_{\star,9}\cdot v_{\star,0}^{-2}\cdot
\rho_{\star,5}
\label{eq:en_loss_num}
\end{equation}
with $\rho_{\star,5}:=\rho_{\star}/(10^{5}\,{\rm g\,cm^{-3}})$. The
energy loss with the four different object types discussed here are
listed in table \ref{results:tab}.
%%%%%%%%%%%%%%%%%%%%%%%%%%%%%%%%%%%%%%%%
\section{PBH captures\label{captures:sec}}
%%%%%%%%%%%%%%%%%%%%%%%%%%%%%%%%%%%%%%%%
A capture of a PBH in a star can by itself lead to much more
significant effects than a single passage. It provides the
possibility of continuous radiation or it can even cause the collapse
of the star.
The PBH is captured by a stellar object, if the following conditions
are met:
\begin{enumerate}
\item The energy loss must be larger than the initial energy,
\begin{equation}
E_{\rm loss}> E_{\rm ini}=\frac{1}{2}\,m_{pbh}\,v_{0}^{2}\,,
\end{equation}
\item The point of return $R$ of the PBH must be smaller than half the
  distance to the next star, $R<1/2\cdot R_{dist}$ - otherwise, the PBH would randomly pass
  through a number of different stars. While this would also result in
  a diffuse, but relatively weak signal, we investigate the case of an
  actual capture. Here, we assume an average distance of $R_{dist}\approx 1$~pc between
  stars.
\end{enumerate}
Using energy conservation between the point when the PBH exits the
star and its point of return to the star
\begin{equation}
\frac{1}{2},m_{pbh}\,v_{0}^{2}-E_{\rm
  loss}=-G\frac{m_{pbh}\,M_{\star}}{R}>-G\frac{m_{pbh}\,M_{\star}}{\frac{1}{2}\,R_{dist}}\,.
\label{energy_conserv:equ}
\end{equation}
The inequality corresponds to condition (2),
\mbox{$R<1/2\,R_{dist}$}. 
We can now insert Equ.\ (\ref{ndot:equ2}) and
(\ref{eq:en_loss_num}) in Equ.\ (\ref{energy_conserv:equ}):
\begin{eqnarray}
\frac{1}{2},m_{pbh}\,v_{0}^{2}&-&2\cdot 10^{27}\,{\rm erg}\,\cdot m_{pbh,20}^{2}\cdot R_{\star,9}\cdot v_{\star,0}^{-2}\cdot
\rho_{\star,5}\nonumber\\&>&-G\frac{m_{pbh}\,M_{\star}}{\frac{1}{2}\,R_{dist}}\,.
\end{eqnarray}
It can now be solved to give a lower limit
to the PBH mass that can be captured,
\begin{eqnarray}
m_{pbh}&>&m_{pbh}^{capture}=1.2\cdot 10^{27}\,\mbox{g}\cdot\nonumber \\
&\cdot&\left(R_{\star,9}\cdot \rho_{\star,5} \right)^{-1}\cdot
\left(\frac{v_{\star}}{2.2\cdot 10^{7}\,{\rm cm/s}}\right)^{2}\,.
\end{eqnarray}
Here, we denote the lowest possible mass to be captured as
$m_{pbh}^{capture}$. We list the actual values for the minimum capture
masses in table \ref{results:tab}.
Our result implies that PBHs in the unexplored mass range,
$10^{15}$~g~$<m_{pbh}<10^{26}$~g, are typically not captured by
stars. The mass range of $10^{28}$~g~$<m_{pbh}<10^{32}$~g, on the
other hand, can potentially be explored with PBH captures, since
limits in this mass range still allow for a PBH contribution of the order
of 10\% to the dark matter. 
%%%%%%%%%%%%%%%%%%%%%%%%%%%%%%%%%%%%%%%%
\section{Results and conclusions\label{conclusions:sec}}
%%%%%%%%%%%%%%%%%%%%%%%%%%%%%%%%%%%%%%%%
We considered two processes for a possible emission from
PBH interactions with stellar objects: a large number of single
passages or a smaller number of captures. In this final section, we
investigate if it is feasible to detect such signatures.
\subsection{Detectability of a single passage}
\begin{figure}[!t]
\centering
\includegraphics[width=\linewidth]{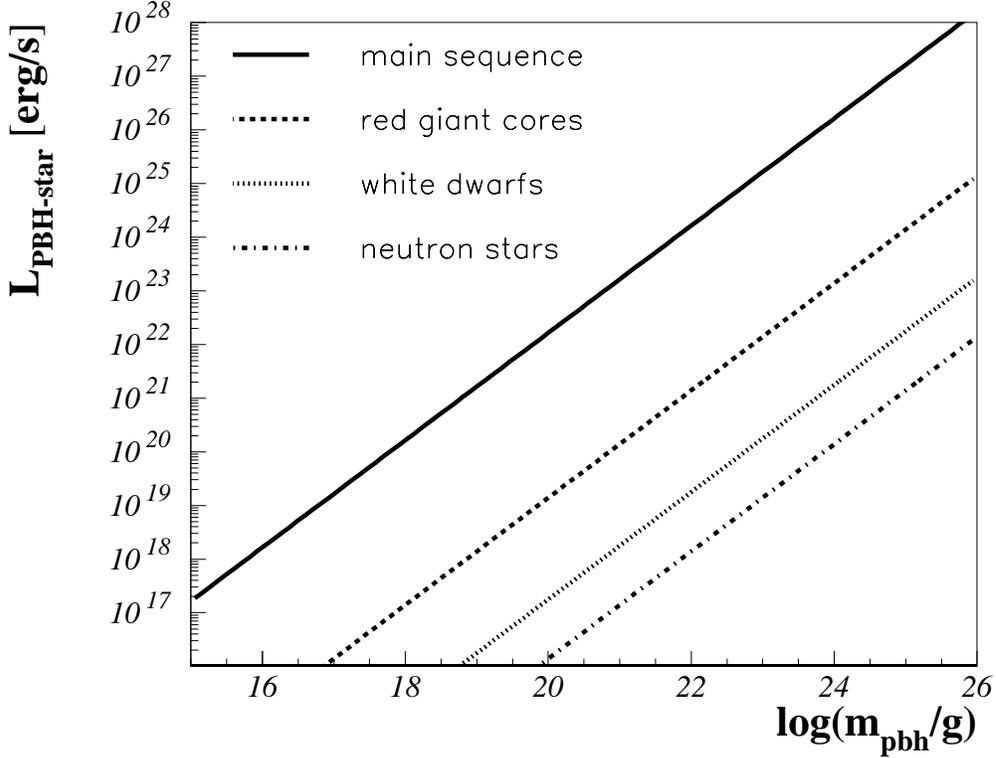}
\caption{Total luminosity from PBH interactions with stellar objects. Solid line: MS$\star$; dashed line: RGC;
  dotted line: WD; dashed line: N$\star$.}
\label{total_luminosity:fig}
\end{figure}

In Section \ref{interaction:sec}, we showed that the event rate of PBH
interactions with stellar object scales as $\dot{n}\propto
m_{pbh}^{-1}$, implying that PBH passages are very frequent at  low
masses and decrease towards high masses. On the other hand, the total energy
loss scales as $m_{pbh}^{2}$, as shown in Section
\ref{energy_loss:sec}. This means that the most powerful events
happen for large mass PBHs. 
We can combine the two properties into a total
luminosity arising from PBH interactions with stellar objects in the
Galaxy:
\begin{equation}
L=\dot{n}\cdot E_{\rm loss}\,.
\end{equation}
For the different object classes, we get numerical values of
\begin{equation}
L=m_{pbh,20}\cdot \left\{\begin{array}{lll}2\cdot 10^{22}\,{\rm erg/s}&&\mbox{MS$\star$}\\
1\cdot 10^{19}\,{\rm erg/s}&&\mbox{RGC}\\
2\cdot 10^{17}\,{\rm erg/s}&&\mbox{WD}\\
1\cdot 10^{16}\,{\rm erg/s}&&\mbox{N$\star$}\end{array}\right.\,.
\end{equation}
This calculation is based on the assumption that PBHs can be the
dominant source of the dark matter. This implies that it can only be
valid in the unexplored region between
$10^{15}$~g~$<m_{pbh}<10^{26}$~g. Figure \ref{total_luminosity:fig}
shows the total luminosity versus PBH mass in the interesting mass
range. The maximum luminosity comes from PBHs with a mass of
$10^{26}$~g, interacting with main
sequence stars. Even for this optimal case, the luminosity is less
than $2\cdot 10^{28}$~erg/s, i.e.\ far below solar
luminosities. Thus, single passages of PBHs through stellar objects
cannot be observed by any detector, since the luminosity lies far
below what is observed from classical radiation processes in the Galaxy.
%===============================================
\subsection{Probability of PBH captures}
\begin{figure}[!t]
\centering
\includegraphics[width=\linewidth]{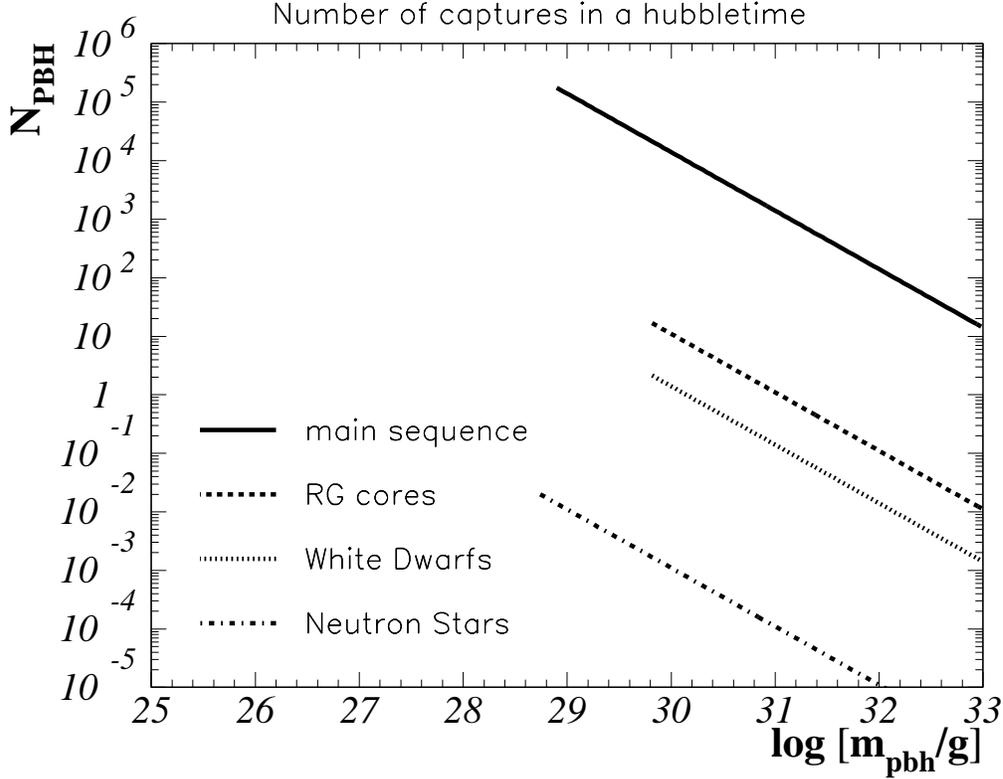}
\caption{Number of captures in a Hubble time, $t_H=10^{10}$~yr, for
  different source classes. Solid line: MS$\star$; dashed line: RGC;
  dotted line: WD; dashed line: N$\star$.}
\label{events_in_hubbletime:fig}
\end{figure}

%===============================================
In Section \ref{captures:sec}, it was shown that captures of PBHs are
possible for masses above $\sim 10^{28}$~g. The question is again, if
these events are frequent enough to lead to detectable effects. Here,
we discuss the mass region $10^{28}$~g~$<m_{pbh}<10^{32}$~g, as PBHs
can still contribute with 10\% to the dark matter. Thus, we use
$\eta=0.1$ in the following calculations. Using Equ.\ (\ref{ndot:equ2}), we determine the number of
interactions between PBHs and stellar objects in a Hubble time,
$t_H = 10^{10}$~yr,
\begin{eqnarray}
N_{pbh}&=&\dot{n}(\eta=0.1)\cdot t_H \\
&=& 10^{6}\cdot N_{\star,9}\cdot
v_{\star,0}\cdot R_{\star,9}^{2}\cdot m_{pbh,20}^{-1}\,.
\end{eqnarray}
The best capture rate is at the lowest mass to be captured,
$m_{pbh}^{capture}$. The results for the different source classes are
listed in table \ref{results:tab}. Figure
\ref{events_in_hubbletime:fig} shows the number of captures versus the
PBH mass. The number of captured objects in red giant cores, white dwarfs and
neutron stars are of the order of 1 in a Hubble time or
smaller. Clearly, this is a negligible effect. Main sequence stars, on
the other hand, can capture up to $2\cdot 10^4$ PBHs in a Hubble time,
which can lead to significant effects like enhanced high-energy
radiation from those stars or even the collapse of a star due to the
influence of the PBH. Calculations on the energy loss and further
influence of the PBH on the stars are in preparation. As we have used
the most optimistic assumptions for our calculations so far, a more
detailed investigation of the number of captured PBHs is necessary as
well. With a detailed description of the energy loss and the event
rate, captures of PBHs in main sequence stars may lead to limits of
their contribution to the dark matter, or even to the detection of
their existence.  Apart from using our own Galaxy, dwarf elliptical galaxies are a good target for
observation, as they are nearby, well-defined observation targets with a low
background rate. 
%\subsection{Summary}
\begin{table}[!h]
  \caption{Results for the energy loss and event rate in a stellar object for a PBH with a mass
  of $10^{20}$~g. The energy loss scales as $E_{\rm loss} \propto m_{pbh,20}^{2}$, while
  the event rate behaves as $\dot{n}\propto m_{pbh,20}^{-1}$. The
  luminosity $L$ is the product of the two. We
  further give the lower limit for masses of PBHs to be captured and the
  expected number of captures.}
  \label{results:tab}
  \centering
  \begin{tabular}{|l|l|l|l|l|}
  \hline
   &MS$\star$  & RGC & WD & N$\star$\\
   \hline 
   $\dot{n} [$yr$^{-1}]$
& & & & \\
  $(m_{pbh}=10^{20}$~g$)$ &$1.3\cdot10^{4}$&$11$&$1.4$&$1.1\cdot 10^{-3}$ \\
   $E_{\rm loss}$ [erg]&&&&\\ \hline
$(m_{pbh}=10^{20}$~g$)$&$4\cdot 10^{25}$&$4\cdot
   10^{25}$ &$4\cdot 10^{24}$ &$4\cdot 10^{26}$\\ \hline
   $L$ [erg/s]&&&&\\
$(m_{pbh}=10^{20}$~g$)$&$2\cdot 10^{22}$&$1\cdot 10^{19}$ &$2\cdot 10^{18}$ &$1\cdot 10^{16}$ \\\hline
   $m_{pbh}^{capture}$ [g]&$8\cdot 10^{28}$&$6\cdot 10^{28}$ &$6\cdot
   10^{29}$ &$6\cdot 10^{27}$ \\\hline
$N_{pbh}$&&&&\\
  $(m_{pbh}=m_{pbh}^{capture})$ &$2\cdot10^{4}$&$2$&$0.2$&$2\cdot 10^{-3}$ \\
  \hline
  \end{tabular}
  \end{table}
%=============================================
\subsection*{Acknowledgments}
%=============================================
JKB acknowledges the support by the DFG grant BE-3714/3-1. MAA acknowledges the support from the Polish Ministry of Science, grant
N203 0093/1466, and from
the Swedish Research Council, grant VR Dnr 621-2006-3288. Support for
the work of PLB has come from the AUGER membership and theory grant 05 CU 5PD 1/2 via DESY/BMBF and VIHKOS.
  
\end{document}